\documentclass{article}
\usepackage{amsfonts}
\usepackage{graphicx}
\usepackage[mathlines]{lineno}
\usepackage{amsmath,amsthm}

\newtheorem*{notation}{Notation}

\newtheorem*{proposition}{Proposition}
\usepackage[colorlinks=true]{hyperref}
\hypersetup{urlcolor=blue, citecolor=red}
\newcommand*\patchAmsMathEnvironmentForLineno[1]{  \expandafter \let \csname old#1\expandafter \endcsname \csname#1\endcsname
  \expandafter \let \csname oldend#1\expandafter \endcsname \csname
end#1\endcsname
  \renewenvironment{#1}     {\linenomath \csname old#1\endcsname}     {\csname oldend#1\endcsname \endlinenomath}}
\newcommand*\patchBothAmsMathEnvironmentsForLineno[1]{  \patchAmsMathEnvironmentForLineno{#1}  \patchAmsMathEnvironmentForLineno{#1*}}
\patchBothAmsMathEnvironmentsForLineno{equation}
\patchBothAmsMathEnvironmentsForLineno{align}
\patchBothAmsMathEnvironmentsForLineno{flalign}
\patchBothAmsMathEnvironmentsForLineno{alignat}
\patchBothAmsMathEnvironmentsForLineno{gather}
\patchBothAmsMathEnvironmentsForLineno{multline}
\usepackage{newtxtext,newtxmath}
\begin{document}
\title{\vspace{-1in}%
%\parbox{\linewidth}{\footnotesize\noindent
%\textbf{Applied Mathematics E-Notes, xx(20xx), xx-xx} \copyright \hfill ISSN 1607-2510
%\newline
%Available free at mirror sites of http://www.math.nthu.edu.tw/$\sim$amen/} 
%\vspace{0pt} \\
Involuntary unemployment in overlapping generations model due to instability of the economy\thanks{%
Mathematics Subject Classifications: 91B64.}}
\date{}
%\date{{\small Received 20 July 2000}}
\author{Yasuhito Tanaka\thanks{%
Faculty of Economics, Doshisha University, Kamigyo-ku, Kyoto, 602-8580, Japan.}}
\maketitle

\begin{abstract}
The existence of involuntary unemployment advocated by J. M. Keynes is a very important problem of the modern economic theory. Using a three-generations overlapping generations model, we show that the existence of involuntary unemployment is due to the instability of the economy. Instability of the economy is the instability of the difference equation about the equilibrium price around the full-employment equilibrium, which means that a fall in the nominal wage rate caused by the presence of involuntary unemployment further reduces employment. This instability is due to the negative real balance effect that occurs when consumers' net savings (the difference between savings and pensions) are smaller than their debt multiplied by the marginal propensity to consume from childhood consumption.
\end{abstract}

\section{Introduction}

The existence of involuntary unemployment advocated by J. M. Keynes is a very important problem of the modern economic theory. It is a phenomenon that workers are willing to work at the market wage or just below but are prevented by factors beyond their control, mainly, deficiency of aggregate demand. \cite{umada1} derived an upward-sloping labor demand curve from the mark-up principle for firms, and argued that such an upward-sloping labor demand curve leads to the existence of involuntary unemployment without wage rigidity\footnote{\cite{lav} presented a similar analysis.}. But his model of firm behavior is ad-hoc. \cite{otaki2} assumes indivisibility  of labor supply, and has shown the existence of involuntary unemployment using efficient wage bargaining according to \cite{solow1}. The arguments of this paper do not depend on bargaining. As discussed by \cite{otaki-agtel} and \cite{otakib} (Theorem 2.3), if labor supply is divisible and very small, no unemployment exists\footnote{If labor supply is indivisible, it may be 1 or 0. On the other hand, in contrast if it is divisible, it takes a real value between 0 and 1. About indivisible labor supply also please see \cite{hansen1}. In \cite{singa1}, \cite{TEL1} and \cite{eb20-1} involuntary unemployment under indivisible labor supply is analyzed.}. However, we show that even if labor supply is divisible, there may exist involuntary unemployment. 

In this paper we consider consumers' utility maximization and firms' profit maximization in an overlapping generations (OLG) model under monopolistic competition according to \cite{otaki1}, \cite{otaki2}, \cite{otaki-agtel} and \cite{otakib}. We extend Otaki's model to a three-generations OLG model with a childhood period and pay-as-you-go pension system for the older generation consumers. We show that the existence of involuntary unemployment is due to the instability of the economy. Instability of the economy is the instability of the difference equation about the equilibrium price around the full-employment equilibrium, which means that a fall in the nominal wage rate caused by the presence of involuntary unemployment reduces employment. In the next section we explain the model and show the existence of involuntary unemployment when aggregate demand is insufficient. In Section 3 we will show the following results.
\begin{enumerate}
\item If the net savings (the difference between savings and pensions) is greater than debts (due to consumption in childhood period) of consumers, then the positive real balance effect kicks in, and involuntary unemployment will spontaneously dissipate because the decline in nominal wages and prices due to unemployment reduces unemployment.
\item If the net savings is smaller than debts of consumers, then the negative real balance effect kicks in, and involuntary unemployment does not spontaneously dissipate because the decline in the nominal wage and prices due to unemployment further increases unemployment.
\end{enumerate}

\section{The model and analysis}

\subsection{Consumers' utility maximization}

 We consider a three-periods (0: childhood, 1: younger or working, and 2: older or retired) OLG model under monopolistic competition. It is a re-arrangement and an extension of the model put forth by \cite{otaki1}, \cite{otaki2}, and \cite{otakib}. The structure of our model is as follows.  
\begin{enumerate}
\item There is one factor of production, labor, and there is a continuum of perishable goods indexed by $z\in [0, 1]$. Good $z$ is monopolistically produced by firm $z$ with constant returns to scale technology. 
\item Consumers consume the goods during the childhood period (Period 0). This consumption is covered by borrowing money from (employed) consumers of the younger generation and/or scholarships. They must repay these debts in their Period 1. However, unemployed consumers cannot repay their own debts. Therefore, we assume that unemployed consumers receive unemployment benefits from the government, which are covered by taxes on employed consumers of the younger generation.

\item During Period 1, consumers supply $l$ units of labor, repay the debts and save money for their consumption in Period 2. They also pay taxes for the pay-as-you go pension system for the older generation. 

\item During Period 2, consumers consume the goods using their savings carried over from their Period 1 earnings, and receive the pay-as-you go pension, which is a lump-sum payment. It is covered by taxes on employed consumers of the younger generation.

\item Consumers determine their consumptions in Periods 1 and 2 and the labor supply at the beginning of Period 1. We assume that their consumption during the childhood period is constant. 
\end{enumerate}

Further we make the following assumptions

\begin{description}
	\item[\textbf{Ownership of the firms}] Each consumer inherits ownership of the firms from the previous generation. Corporate profits are distributed equally to consumers.
	\item[\textbf{Zero interest rate}] We assume zero interest rate, and that repayment of the debts of consumers in their childhood period is assured. Consumer borrowing in childhood period is constant. If the savings of consumers in the younger period are insufficient for the borrowing, the government lends the scholarship to consumers in the childhood period. Consumers in the younger period are indifferent between lending money to childhood period consumers and savings by money.

%Due to the existence of pay-as-you-go pension the savings is likely to be insufficient for borrowing when consumption by consumers in the childhood period is not so small.
\end{description}

\begin{notation}
We use the following notation. 

\begin{tabular}{l}
$C^e_i$: consumption basket of an employed consumer in Period $i,\ i=1,2$.\\
$C^u_i$: consumption basket of an unemployed consumer in Period $i,\ i=1,2$.\\
$c_i^{e}(z)$: consumption of good $z$ of an employed consumer in Period $i,\ i=1,2$.\\
$c_i^{u}(z)$: consumption of good $z$ of an unemployed consumer in Period $i,\ i=1,2$.\\
$D$: consumption basket of an individual in the childhood period, which is constant.\\
$P_i$: the price of consumption basket in Period $i,\ i=1,2$.\\
$p_i(z)$: the price of good $z$ in Period $i,\ i=1,2$.\\
$\rho=\frac{P_2}{P_1}$: (expected) inflation rate (plus one).\\
$W$: nominal wage rate.\\
$R$: unemployment benefit for an unemployed individual. $R=D$.\\
$\hat{D}$: consumption basket in the childhood period of a next generation consumer.\\
$Q$: pay-as-you-go pension for an individual of the older generation.\\
$\Theta$: tax payment by an employed individual for the unemployment benefit.\\
$\hat{Q}$: pay-as-you-go pension for an individual of the younger generation when he retires.\\
$\Psi$: tax payment by an employed individual for the pay-as-you-go pension.\\
$\Pi$: profits of firms which are equally distributed to each consumer.\\
$l$: labor supply of an individual.\\
$\Gamma(l)$: disutility function of labor, which is increasing and convex.
\end{tabular}

\begin{tabular}{l}
$L$: total employment.\\
$L_f$: population of labor or employment in the full-employment state. \\
$y$: labor productivity, which is constant.
\end{tabular}
\end{notation}

We assume that the population $L_f$ is constant. We also assume that the nominal wage rate is constant in this section,. We examine the effects of a change in the nominal wage rate in Section 3.

We consider a two-step method to solve utility maximization of consumers such that:
\begin{enumerate}
\item Employed and unemployed consumers maximize their utility by determining consumption baskets in Periods 1 and 2 given their income over two periods:
\item Then, they maximize their consumption baskets given the expenditure in each period.
\end{enumerate}

Since the taxes for unemployed consumers' unemployment benefits are paid by employed consumers of the same generation, $D(=R)$ and $\Theta$ satisfy $D(L_f-L)=L\Theta$. It means 
\begin{equation*}
L(D+\Theta)=L_fD.%\label{e1}
\end{equation*}
The price index of the consumption basket in Period 0 is assumed to be 1. Thus, $D$ is the real value of the consumption in the childhood period of consumers. 

Also, since the taxes for the pay-as-you-go pension system are paid by employed consumers of younger generation, $Q$ and $\Psi$ satisfy the following relationship:
\[L\Psi=L_fQ.\]
The utility function of employed consumers of one generation over three periods is
\[u(C^e_1,C^e_2,D)-\Gamma(l).\]
We assume that $u(\cdot)$ is a homothetic utility function. The utility function of unemployed consumers is
\[u(C^u_1,C^u_2,D).\]
The consumption baskets of employed and unemployed consumers in Period $i$ are
\[C^e_i=\left(\int_0^1c_i^e(z)^{\frac{\sigma-1}{\sigma}}dz\right)^{\frac{\sigma}{\sigma-1}},\ C^u_i=\left(\int_0^1c_i^u(z)^{\frac{\sigma-1}{\sigma}}dz\right)^{\frac{\sigma}{\sigma-1}},\ i=1,2.\]
$\sigma$ is the elasticity of substitution among the goods, and $\sigma>1$.

The price of consumption basket in Period $i$ is
\[P_i=\left(\int_0^1  p_i(z)^{1-\sigma}dz\right)^{\frac{1}{1-\sigma}},\ i=1,2.\]
The budget constraint for an employed consumer is
\[P_1C^e_1+P_2C^e_2=Wl+\Pi-D-\Theta+\hat{Q}-\Psi.\]
The budget constraint for an unemployed consumer is 
\[P_1C^u_1+P_2C^u_2=\Pi-D+R+\hat{Q}=\Pi+\hat{Q}\ (\mathrm{since}\ R=D).\]
Let 
\begin{equation*}
\alpha=\frac{P_1C^e_1}{P_1C^e_1+P_2C^e_2},\ 1-\alpha=\frac{P_2C^e_2}{P_1C^e_1+P_2C^e_2}.%\label{al1}
\end{equation*}
Since the utility functions $u(C^e_1,C^e_2,D)$ and $u(C^u_1,C^u_2,D)$  are homothetic, $\alpha$ is determined by the relative price $\frac{P_2}{P_1}$, and do not depend on the income of the consumers. Therefore, we have
\[\alpha=\frac{P_1C^e_1}{P_1C^e_1+P_2C^e_2}=\frac{P_1C^u_1}{P_1C^u_1+P_2C^u_2},\ 1-\alpha=\frac{P_2C^e_2}{P_1C^e_1+P_2C^e_2}=\frac{P_2C^u_2}{P_1C^u_1+P_2C^u_2}.\]
From the first order conditions and the budget constraints for employed and unemployed consumers we obtain the following demand functions for consumption baskets.
\[C^e_1=\alpha\frac{Wl+\Pi-D-\Theta+\hat{Q}-\Psi}{P_1},\ C^e_2=(1-\alpha)\frac{Wl+\Pi-D-\Theta+\hat{Q}-\Psi}{P_2},\]
\[C^u_1=\alpha\frac{\Pi+\hat{Q}}{P_1},\ C^u_2=(1-\alpha)\frac{\Pi+\hat{Q}}{P_2}.\]
Solving maximization problems in Step 2 by standard calculations, the following demand functions of employed and unemployed consumers are derived.
\begin{equation*}
c^e_1(z)=\left(\frac{p_1(z)}{P_1}\right)^{-\sigma}\frac{\alpha(Wl+\Pi-D-\Theta+\hat{Q}-\Psi)}{P_1},
\end{equation*}
\begin{equation*}
c^e_2(z)=\left(\frac{p_2(z)}{P_2}\right)^{-\sigma}\frac{(1-\alpha)(Wl+\Pi-D-\Theta+\hat{Q}-\Psi)}{P_2},
\end{equation*}
\begin{equation*}
c^u_1(z)=\left(\frac{p_1(z)}{P_1}\right)^{-\sigma}\frac{\alpha(\Pi+\hat{Q})}{P_1},\ c^u_2(z)=\left(\frac{p_2(z)}{P_2}\right)^{-\sigma}\frac{(1-\alpha)(\Pi+\hat{Q})}{P_2}.
\end{equation*}
From these analyses we obtain the indirect utility functions of employed and unemployed consumers as follows:
\begin{align*}
V^e&=u\left(\alpha\frac{Wl+\Pi-D-\Theta+\hat{Q}-\Psi}{P_1},(1-\alpha)\frac{Wl+\Pi-D-\Theta+\hat{Q}-\Psi}{P_2},D\right)-\Gamma(l),
\end{align*}
\begin{align*}
V^u&=u\left(\alpha\frac{\Pi+\hat{Q}}{P_1},(1-\alpha)\frac{\Pi+\hat{Q}}{P_2},D\right).
\end{align*}
Let $\omega=\frac{W}{P_1},\ \rho=\frac{P_2}{P_1}$. Then, since the real value of $D$ in the childhood period is constant, we can write
\begin{equation*}
V^e=\varphi\left(\omega l+\frac{\Pi-D-\Theta+\hat{Q}-\Psi}{P_1}, \rho\right)-\Gamma(l),\ V^u=\varphi\left(\frac{\Pi+\hat{Q}}{P_1}, \rho\right).
\end{equation*}
$\omega$ is the real wage rate. Denote 
\begin{equation*}
I=\omega l+\frac{\Pi-D-\Theta+\hat{Q}-\Psi}{P_1}.%\label{i}
\end{equation*}
The condition for maximization of $V^e$ with respect to $l$ given $\rho$ is
\begin{equation}
\frac{\partial \varphi}{\partial I}\omega-\Gamma'(l)=0,\label{ve}
\end{equation}
where
\[\frac{\partial \varphi}{\partial I}=\alpha\frac{\partial u}{\partial C^e_1}+(1-\alpha)\frac{\partial u}{\partial C^e_2}.\]
Given $P_1$ and $\rho$ the labor supply is a function of $\omega$. From (\ref{ve}) we get
\begin{equation*}
\frac{dl}{d\omega}=\frac{\frac{\partial \varphi}{\partial I}+\frac{\partial^2 \varphi}{\partial I^2}\omega l}{\Gamma''(l)-\frac{\partial^2 \varphi}{\partial I^2}\omega^2}.
\end{equation*}
If $\frac{dl}{d\omega}>0$, the labor supply is increasing with respect to the real wage rate $\omega$. Labor supply $l$ may depend on the employment $L$. We assume that $Ll$ is increasing in $L$.

\subsection{Firms' profit maximization}

Let $d_1(z)$ be the total demand for good $z$ by younger generation consumers in Period 1. Then, 
\begin{align*}
d_1(z)=\left(\frac{p_1(z)}{P_1}\right)^{-\sigma}\frac{\alpha\left(WLl+L_f\Pi-L_fD+L_f\hat{Q}-L_fQ\right)}{P_1}.
\end{align*}
This is the sum of the demand of employed and unemployed consumers. Note that $\hat{Q}$ is the pay-as-you-go pension for younger generation consumers in their Period 2. Similarly, their total demand for good $z$ in Period 2 is written as
\[d_2(z)=\left(\frac{p_2(z)}{P_2}\right)^{-\sigma}\frac{(1-\alpha)\left(WLl+L_f\Pi-L_fD+L_f\hat{Q}-L_fQ\right)}{P_2}.\]
Let $\overline{d_2(z)}$ be the demand for good $z$ by the older generation. Then,
\[\overline{d_2(z)}=\left(\frac{p_1(z)}{P_1}\right)^{-\sigma}\frac{(1-\bar{\alpha})\left(\bar{W}\bar{L}\bar{l}+L_f\bar{\Pi}-L_f\bar{D}+L_fQ-L_f\bar{Q}\right)}{P_1},\]
where $\bar{W}$, $\bar{\Pi}$, $\bar{L}$, $\bar{l}$, $\bar{D}$ and $\bar{Q}$ are the nominal wage rate, the profits of firms, the employment, the individual labor supply, the debt of an individual, and the pay-as-you-go pension, respectively, during the previous period. $\bar{\alpha}$ is the value of $\alpha$ for the older generation. $Q$ is the pay-as-you-go pension for consumers of the older generation themselves. Let 
\begin{equation*}
M=(1-\bar{\alpha})\left(\bar{W}\bar{L}\bar{l}+L_f\bar{\Pi}-L_f\bar{D}+L_fQ-L_f\bar{Q}\right).%\label{m}
\end{equation*}
This is the total savings or the total consumption of the older generation consumers including the pay-as-you-go pensions they receive in their Period 2. It is the planned consumption that is determined in Period 1 of the older generation consumers. \emph{Net savings} is the difference between $M$ and the pay-as-you-go pensions in their Period 2, as follows:
\begin{equation*}
\tilde{M}=M-L_fQ.%\label{m1}
\end{equation*}
Their demand for good $z$ is written as $\left(\frac{p_1(z)}{P_1}\right)^{-\sigma}\frac{M}{P_1}$. Government expenditure constitutes the national income as well as the consumptions of the younger and older generations. It is financed by the tax on the younger generation consumers. Then, the total demand for good $z$ is written as
\begin{equation}
d(z)=\left(\frac{p_1(z)}{P_1}\right)^{-\sigma}\frac{Y}{P_1},\label{dz}
\end{equation}
where $Y$ is the effective demand defined by 
\[Y=\alpha\left(WLl+L_f\Pi-T-L_fD+L_f\hat{Q}-L_fQ\right)+G+L_f\hat{D}+M.\]
Note that $\hat{D}$ is consumption in the childhood period of a next generation consumer. $G$ is the government expenditure, except for the pay-as-you-go pensions, scholarships and unemployment benefits, and $T$ is the tax revenue for the government expenditure. See \cite{otaki1}, \cite{otakib} about this demand function. 
 
Let $L$ and $Ll$ be employment and the ``employment $\times$ labor supply'' of firm $z$. The output of firm $z$ is $Lly$. At the equilibrium $Lly=d(z)$. Then, we have
\[\frac{\partial d(z)}{\partial (Ll)}=y.\]
From (\ref{dz}) $\frac{\partial p_1(z)}{\partial d(z)}=-\frac{p_1(z)}{\sigma d(z)}$. Thus
\[\frac{\partial p_1(z)}{\partial (Ll)}=-\frac{p_1(z)y}{\sigma d(z)}=-\frac{p_1(z)y}{\sigma Lly}.\]
%\[\frac{\partial d(z)}{\partial p_1(z)}=-\sigma \frac{d(z)}{p_1(z)}.\]
The profit of firm $z$ is 
\[\pi(z)=p_1(z)Lly-LlW.\]
The condition for profit maximization is 
\begin{align*}
\frac{\partial \pi(z)}{\partial (Ll)}=&p_1(z)y-Lly\frac{p_1(z)y}{\sigma Lly}-W=p_1(z)y-\frac{p_1(z)y}{\sigma}-W=0.
\end{align*}
Therefore, we obtain
\begin{equation*}
p_1(z)=\frac{1}{(1-\frac{1}{\sigma})y}W=\frac{1}{(1-\mu)y}W,\ \mu=\frac{1}{\sigma}.
\end{equation*}
This means that the real wage rate is
\begin{equation*}
\omega=(1-\mu)y. %\label{real}
\end{equation*}
Since all firms are symmetric, 
\begin{equation}
P_1=p_1(z)=\frac{1}{(1-\mu)y}W.\label{price}
\end{equation}

\subsection{Involuntary unemployment due to instability of the economy}

Consider an economy at Period $t$. The (nominal) aggregate supply of the goods is equal to 
\[W^tL^t+L_f\Pi^t=P_1^tL^tl^ty.\]
The (nominal) aggregate demand is
\begin{align*}
&\alpha\left(W^tL^t+L_f\Pi^t-T^t-L_fD^t+L_f\hat{Q}^t-L_fQ^t\right)+G^t+L_f\hat{D}^t+M^t\\
=&\alpha\left(P_1^tL^tl^ty-T^t-L_fD^t+L_f\hat{Q}^t-L_fQ^t\right)+G^t+L_f\hat{D}^t+M^t.
\end{align*}
The superscript $t$ denotes variables at Period $t$. Since the aggregate demand and supply are equal in the equilibrium, 
\begin{equation*}
P_1^tL^tl^ty=\alpha\left(P_1^tL^tl^ty-T^t-L_fD^t+L_f\hat{Q}^t-L_fQ^t\right)+G^t+L_f\hat{D}^t+M^t.
\end{equation*}
We obtain $L^tl^t$ as follows:
\begin{align}
L^tl^t=&\frac{\alpha\left(-T^t-L_fD^t+L_f\hat{Q}^t-L_fQ^t\right)+G^t+L_f\hat{D}^t+M^t}{(1-\alpha)P_1^ty}.\label{e2}
\end{align}
$L^tl^t$ cannot be larger than $L_fl(L_f)$, where $l(L_f)$ is the labor supply at full-employment. However, it may be strictly smaller than $L_fl(L_f)$. Then, we have $L^t<L_f$ and involuntary umemployment exists. We assume balanced budget $G^{t}=T^{t}$. In the full-employment equilibrium without excess demand $L^tl^t=L_fl(L_f)$, $P_1^{t+1}=P_1^t$, $\hat{Q}^t=Q^t$, $\hat{D}^t=D^t$. Denote the variables in the full-employment equilibrium by a superscript $^*$. Then,
\[L_fl(L_f)=\frac{\alpha\left(-G^*-L_fD^*+L_fQ^*-L_fQ^*\right)+G^*+L_fD^*+M^*}{(1-\alpha)P_1^*y}=\frac{(1-\alpha)(G^*+L_fD^*)+M^*}{(1-\alpha)P_1^*y}.\]
Let us denote the real values of $G^t$, $\hat{D}^t$ and $Q^t$, respectively, by $g$, ${d}$ and $q$. We assume that the real values of these variables are maintained even if the prices change. Then,
\[L_fl(L_f)=\frac{(1-\alpha)(P_1^*g+L_fP_1^*d)+M^*}{(1-\alpha)P_1^*y}\]
This means
\[M^*=(1-\alpha)P_1^*(L_fl(L_f)y-g-L_fd).\]
Suppose that when there exists involuntary unemployment, the nominal wage rate falls. Then, the prices of the goods also fall at the same rate because of the constant returns to scale according to (\ref{price}). This relation is expressed by the following difference equation.
\begin{equation*}
P_1^{t+1}=\gamma \left(L^tl^t-L_fl(L_f)\right)+P_1^t,\ \gamma>0.%\label{ww}
\end{equation*}
Let us denote 
\[P_1^{t+1}=f(P_1^t).\]
We assume $f'(P_1^t)>0$. Since $L_fl(L_f)$ is constant,
\begin{equation}
f'(P_1^t)=\gamma \frac{d L^tl^t}{dP_1^t}+1.\label{st1}
\end{equation}
According to Chap. 4 of \cite{Schreiber} the stability condition for the full-employment equilibrium is
\[f'(P_1^t)<1\ \mathrm{at}\ P_1^t=P_1^*.\]
The total savings or total consumption of the older generation $M^t$ is not constant nor predetermined, but the net savings
\[\tilde{M}^t=M^t-L_fP_1^tq\]
is predetermined. Also note that $\hat{Q}^t=P_1^{t+1}q$ in (\ref{e2}). From (\ref{e2}) with $G^t=T^t=P_1^t g$, $Q^t=P_1^tq$, $\hat{Q}^t=P_1^{t+1}q$ and $\hat{D}^t=P_1^td$,
\begin{align*}
f'(P_1^t)=\gamma\frac{L_fP_1^tq-\alpha\left(-L_fD^t+L_fP_1^{t+1}q\right)-{M}^t}{(1-\alpha)\left(P_1^t\right)^2y}+\gamma\frac{\alpha L_fP_1^tf'(P_1^t)q}{(1-\alpha)\left(P_1^t\right)^2y}+1,
\end{align*}
where $\frac{\partial M^t}{\partial P_1^t}=\frac{\partial L_fP_1^tq}{\partial P_1^t}=L_fq$. At $P_1^t=P_1^*$,
\begin{align*}
f'(P_1^*)=\gamma\frac{(1-\alpha)L_fP_1^*q+\alpha L_fP_1^*d-{M}^*}{(1-\alpha)\left(P_1^*\right)^2y}+\gamma\frac{\alpha L_fP_1^*f'(P_1^*)q}{(1-\alpha)\left(P_1^*\right)^2y}+1,
\end{align*}
Therefore,
\[f'(P_1^*)=\frac{\gamma \left[(1-\alpha)L_fP_1^*q+\alpha L_fP_1^*d-{M}^*\right]+(1-\alpha)\left(P_1^*\right)^2y}{(1-\alpha)\left(P_1^*\right)^2y-\gamma\alpha L_fP_1^*q}=-\frac{\gamma\left(M^*-L_fP_1^*q-\alpha L_fP_1^*d\right)}{(1-\alpha)\left(P_1^*\right)^2y-\gamma\alpha L_fP_1^*q}+1.\]
Then, since $f'(P_1^*)>0$, $f'(P_1^*)<1$ is equivalent to
\[M^*-L_fP_1^*q-\alpha L_fP_1^*d>0.\]
On the other hand, in contrast $f'(P_1^*)>1$ is equivalent to
\begin{equation}
M^*-L_fP_1^*q-\alpha L_fP_1^*d<0.\label{st2}
\end{equation}
From (\ref{st1}), if $f'(P_1^*)>1$, then $\left.\frac{d L^tl^t}{dP_1^t}\right|_{P_1^t=P_1^{t+1}=P_1^*}>0$, which implies that a fall in the price of the goods decreases the employment, and involuntary unemployment won't go away naturally. This and (\ref{st2}) mean that due to the fact that consumer debt multiplied by the marginal propensity to consume is greater than net savings, the negative real balance effect works.

We have shown the following results.
\begin{proposition}
\begin{enumerate}
\item If the net savings (the difference between savings and pensions) is greater than debts (due to consumption in childhood period) of consumers, then the positive real balance effect kicks in, and involuntary unemployment will spontaneously dissipate because the decline in nominal wages and prices due to unemployment reduces unemployment.
\item If the net savings is smaller than debts of consumers, then the negative real balance effect kicks in, and involuntary unemployment does not spontaneously dissipate because the decline in the nominal wage and prices due to unemployment further increases unemployment.
\end{enumerate}
\end{proposition}

\begin{description}
	\item[Acknowledgment] 

This work was supported by the Japan Society for the Promotion of Science KAKENHI (Grant Number 18K01594).
\end{description}

\end{document}